\documentclass{article} 
\usepackage{graphicx}        % standard LaTeX graphics tool
\begin{document}
\noindent
{\huge \bf Statistical Physics for Humanities: }

\medskip
\noindent
{\huge \bf A Tutorial}

\bigskip
\noindent
Dietrich Stauffer

\bigskip
\noindent
{\small 
The image of physics is connected with simple ``mechanical'' deterministic 
events: that an apple always falls down, that force equals mass times 
acceleleration. Indeed, applications of such concept to social or historical
problems go back two centuries (population growth and stabilisation, by
Malthus and by Verhulst) and use ``differential equations'', as recently
revierwed by Vitanov and Ausloos [2011].
However, since even today's computers cannot follow the motion of all air 
molecules within one cubic centimeter, the probabilistic approach has become
fashionable since Ludwig Boltzmann invented Statistical Physics in the 19th 
century. Computer simulations in Statistical Physics deal with single particles,
a method called
agent-based modelling in fields which adopted it later. Particularly simple 
are binary models where each particle has only two choices, called spin up and
spin down by physicists, bit zero and bit one by computer scientists, and voters
for the Republicans or for the Democrats in American politics (where one human 
is simulated as one particle). Neighbouring particles may influence each other,
and the Ising model of 1925 is the best-studied example of such models. This 
text will explain to the reader how to program the Ising model on a square 
lattice (in Fortran language); starting from there the readers can build their
own computer programs. Some applications of Statistical Physics outside the
natural sciences will be listed.}

\section {Introduction}

{\em Learning by Doing} is the intention of this tutorial: readers should learn
how to construct their own models and to program them, not learn about the 
great works of the author [Stauffer et al 2006] and the lesser works of his 
competitors [Billari et al 2006]. 

\noindent
Already Empedokles is reported to have 25 centuries ago compared humans to
fluids: Some are easy to mix, like wine and water; and some, like oil
and water, refuse to mix. Newspapers can give you recent human examples, and 
physicists and others have taken up the challenge to study selected problems 
of history and other humanities [Castellano et al 2009] with methods similar to
physics. In the opposite direction, Prados [2009] uses the physics dream of
``unified field theory'' to describe his history of the Vietnam war. 

The next section will recommend ways how to construct models, the following
one how to program a simple Ising model on a square lattice, and a concluding
section will list some applications. An appendix will introduce the Fortran
programming language. 

\section{Model Building}

\subsection{What is a Model?}

\noindent 
``Models'' in physics and in this tutorial usually deal with the single elements
of a system and how their interactions produce the behaviour of the whole
system. Outside of physics, a model may just be any mathematical law 
approximating reality. Thus the statement that human adult mortality increases
exponentially with age is often called the Gompertz model by demographers but the
Gompertz rule by physicists; the latter ones use the Penna model of individuals
undergoing genetic mutations and Darwinian selection to simulate a large 
population perhaps obeying Gompertz [Stauffer et al 2006].  

\subsection{Binary versus more complicated models}

\noindent
If no previous work on a general model is known, I recommend to start with 
binary variables  where each particle has only two choices, called spin up and
spin down by physicists, bit zero and bit one by computer scientists, occupied
or empty for percolation, and voters
for the Republicans or for the Democrats in American politics (where one human 
is simulated as one particle). Of course reality is more complicated, but we 
want to {\em understand} reality: Does it agree with the simplest possible 
model? (Simulations for pilot training etc. are different [Bridson and Batty
2010]). Thus we follow the opinion of Albert Einstein that a model should be as 
simple as possible, but not simpler. If you want to simulate traffic jams 
[Chowdhury et al. 2000] in cities, the colour of the cars is quite irrelevant, 
but for visibility in the dark the colour matters.

Once the binary case has been studied, one can go to more than two choices.
If three groups are fighting each other, obviously three choices
are needed in a model [Lim 2007]. Even in a two-party political system like
the USA, other candidates were important in Florida for the US presidential
elections of 2000. The opinions of people [Malarz et al 2011] are in reality 
continuous and can be modelled by one or several real numbers between zero and 
unity, or 
between minus infinity and plus infinity. Nevertheless it is standard practice
in opinion polls to allow only a few choices like full agreement, partial
agreement, neutrality, partial disagreement, full disagreement. And in elections
one can only vote among the discrete number of candidates or parties which
are on the ballot. In the Kosovo opinion of the International Court of Justice
(July 22, 2010) one judge criticised the binary tradition of either legal or 
illegal, stating that tolerable is in between; nevertheless the court majority
stayed with the binary logic of not illegal.

Physicists like to call the binary variables ``spins'' but readers from
outside physics should refrain from studying the quite complicated spin
concept in quantum mechanics. Spins are simply up and down (1 or 0; 1 or --1).
Similary, don't be deterred if physicist talk about a Hamiltonian; in most
cases this is just the energy from high school. 

\subsection{Humans are neither Spins nor Atoms}

\noindent
Of course, that is true, but it does not exclude that humans are modelled like
spins. Modern medicine blurred the boundary between life and death; nevertheless
we usually talk about people having been born in a certain year, and having died
in another year, as if they were binary up-down variables. Reasons for death
are complicated and I don't even know mine yet; nevertheless demographers like
Gompertz estimated probabilities for dying at some age. Such probabilities are
rather useless for predicting the death of one individual, but averaged over 
many people they may give quite accurate results. When I throw one coin I do not
know how it will fall; when I throw thousand coins, usually about half of them 
fall on one side and the others on the other side (law of large numbers). If I 
throw 1000 coins and all show ``head'', most likely I cheated. Thus to
simulate one person's opinion and decisions on a computer does not seem to 
be realistic; to do the same for millions of people may give good average 
properties, like the number of deaths at the age of 80 to 81 years, or the
fraction of voters selecting the parties in an upcoming election. The whole 
insurance industry is based on this law of large numbers. Humans are
not spins but many humans together might be studied well by spin models.

\subsection{Deterministic or Statistical ?}

\noindent
Non-physicists often believe that physics deals with deterministic rules: The 
apple falls down from the tree and not up; force equals mass times acceleration;
etc. (Or they have heard of quantum-mechanical probability and apply that to 
large systems where such quantum effects should be negligibly small.) In this
sense the cause of World War I was seen as a consequence of the arms race modelled 
by deterministic differential equations for averages [Richardson 1935].  And the
decay of empires was described [Geiss 2008] as starting at the geographical
periphery, since the influence from the center decreases towards zero with 
increasing distance, just as the gravitational force between the sun and its
planets; see also [Diamond 1997, epilog]. 

Because of its historical importance let us look into Richardson's papers of
1935: Two opposing (groups of) nations change their preparedness $x$ for war
because of three reasons: 1) the war preparedness of the other side; 2) fatigue
and expense; 3) dissatisfaction with existing peace treaties. Reasons 1 and 3
increase and reason 2 decreases the war preparedness; reason 1 is proportional
to the $x$ of the opponent, reason 2 to the own $x$, and reason 3 independent of
$x_1$ and $x_2$. Even complete disarmament ($x_1=x_2=0$) at one moment does not 
help if there is dissatisfaction with the existing peace. And if reason 1 is 
stronger than reason 2 for both sides, both $x$ increase exponentially with 
time towards infinity. His second paper details the mathematical solutions for
these linear coupled inhomogenous differential equations.  

More recently, the widespread use of computers shifted the emphasis to more 
realistic {\em probabilistic} models, using random numbers to simulate the 
throwing of coins or other statistical methods. This ``Statistical Physics'' 
and a simple example are the subject of the next section. 

\section{Statistical Physics and the Ising Model}

\subsection{Boltzmann Distribution}

\noindent
No present computer can simulate the motion of all air molecules in a cubic 
centimeter. Fortunately, Ludwig Boltzmann about 150 years ago invented a 
simple rule. The molecules move at temperature $T$ with a velocity which can 
change all the time
but follows a statistical distribution: The probability for a velocity $v$ is
proportional to exp($-E/T$), where $E$ is the kinetic energy of the molecule
due to its velocity. The same principle applied to a binary choice, where 
a particle can be in two states A and B with energies $E_A$ and $E_B$ means 
that the two probabilities are

$$p_A = \frac{1}{Z} \exp(-E_A/T) ; \quad p_B = \frac{1}{Z} \exp(-E_B/T) ; \eqno (1a)$$ 
$$Z = \exp(-E_A/T) + \exp(-E_B/T) \eqno (1b) $$
since the sum over all probabilities must be unity. More generally, a 
configuration with energy $E$ is in thermal equilibrium found with probability

$$ p = \frac{1}{Z} \exp(-E/T) ; \quad Z = \sum \exp(-E/T) \eqno (2) $$
where the sum runs over all possible states of the system. $Z$ is 
called the partition function, one of the rare cases where the German word 
for it, Zu-standssumme = sum over states, is clearer and shorter. The 
temperature $T$ is measured neither in Celsius (centigrade) nor Fahrenheit but
$T=0$ at the absolute zero temperature (about --273 Celsius below the freezing
temperature of water) and moreover is measured in energy units. (If $T$ is 
measured in Kelvin, the corresponding energy is $k_BT$ where $k_B$ is the 
Boltzmann constant and set to unity in the present tutorial.) The function
exp$(x)$ is the exponential function, also written as $e^x$, which for 
integer $x$ means the product of $x$ factors $e \simeq 2.71828$; $e^x = 
2^{x/0.69315} = 10^{0.4343x}$.

Eqs.(1,2) can be regarded as axioms on which Statistical Physics is built,
like the Parallel Axiom of Euclidean Geometry, but in some cases they can be
derived from other principles. Humanities are allowed to use these ideas of
Boltzmann since history institutes were named after him.

\subsection{Ising Model}

\noindent
In 1925, Ernst Ising (born in the heart of the city this author lives in) finished 
his doctoral dissertation on a model for ferromagnetism, which became famous
two decades later and was shown to apply to liquid-vapour transitions half 
a century later. We assume that each site of a lattice (e.g. a square lattice 
where each site $i$ has four neighbours: clockwise  up, right, down, left)
carries a spin $S_i=\pm 1$ (up or down). Neighbouring spins ``want'' to be 
parallel, i.e. they have an energy $-J$ if they are in the same state and an 
energy $+J$ if they are in the two different states. Moreover, a ``magnetic''
field $H$ between minus infinity and plus infinity (also called $B$) tries to 
orient the spins in its own direction. The total energy then is 

$$E = -J \sum_{<ij>} S_i S_j  - H \sum_i S_i         \eqno (3)$$
where the first sum goes over all ordered pairs of neighbor sites $i$ and $j$.
Thus the ``bond'' between sites $A$ and $B$ appears only once in this sum,
and not twice (for $i=A, j=B$ as well as for $i=B, j=A$). The second sum 
runs over all sites of the system. Thus $2J$ is the 
energy to break one bond, and $2H$ is the energy to flip a spin from the 
direction of the field into the opposite direction. As discussed before in 
the Boltzmann subsection, the higher the energy $E$ is the lower is the 
probability to observe this spin configuration; at infinitely high temperatures
$T$ all configurations are equally probable; at $T=0$ all spins must be parallel
to each other and to the field $H$ in equilibrium. The ``magnetisation'' $M$ is
the number of up spins minus the number of down spins,

$$M = \sum_i S_i \quad . \eqno (4)$$
Computer simulations of this Ising model will be described in the appendix.

Applied to human beings, this Ising model could represent two possible
opinions in a population; everybody tries to convince the neighbours of the 
own opinion ($J$), and in addition the government $H$ may try to convince
the whole population of its own opinion. The temperature then gives the 
tendency of the individuals not to think like the majority of their neighbours
and the government. Zero temperature thus means complete conformity, and 
infinite temperature completely random opinions. 

Theories with paper and pencil in two dimensions as well as computer simulations
give $M(T,H)$. In particular, for $H=0$ and in more than one dimension, the
equilibrium magnetisation $M = \pm M_0$ is a non-zero spontaneous magnetisation
for $T < T_c$ and is zero for $T \ge T_c$ where $T_c$ is the critical or Curie
temperature (named after Pierre Curie, not his more famous wife Marie Curie).
On the square lattice, $T_c/J \simeq 2.27$ is known exactly, in three dimensions
only numerically; in one dimension there is no transition to a spontaneous 
magnetisation, as Ernst Ising had shown: $T_c = 0$.

The above model obeys Isaac Newton's law $actio = - reactio$: The sun attracts
the earth with the same force as the earth attracts the sun, only in opposite
direction. Human relations can also be unsymmetrical: He loves her but she does
not love him. Then the bond between sites $i$ and $j$ may be directed instead 
of the usual case of an undirected bond. For example, if $i$ influences $j$ 
but $j$ does not influence $i$, flipping the spin at $j$ from 
parallel ($S_j=+S_i$) to antiparallel ($S_j=-S_i)$ to the spin at $i$ can cost
an energy $2J$ while flipping $S_i$ at constant $S_j$ costs nothing. In this 
case no unique energy $E(S_i,S_j)$ is defined for this spin pair and thus such 
models have been much less studied in the physics literature. (In this example,
one could gain a lot of energy from nothing by the cyclic process of flipping
$j$ from antiparallel to parallel, gaining energy $2J$, then flipping $i$ from 
parallel to antiparallel, costing nothing, then flipping $j$ again and so on.
Such a perpetuum mobile does not exist in physics.) 

Outside physics, of course, one can commit such crimes against energy 
conservation and forget all probabilities proportional to exp($-E/T$). Instead
one can assume arbitrary probabilities, as long as their sum equals one. For
example, if an element has three possible states A, B and C, then one may assume
that with probability $p$ an A becomes B, a B becomes C, and a C becomes A; 
with probability $1-p$ the element does not change, and there is no backward
process from C to B to A to C. Then one has a circular perpetuum mobile, which
may be realistic for some social processes. Physicists sometimes distinguish
between dynamics (when the changes are determined fully by energy or force) and
kinetics (when additional assumptions, like the probabilites of eqs.(1)
are made); probabilities independent of energy/force are then kinetics, as is
most of the material described here. And usually to find a stationary or
static equilibrium, one has to wait for many non-equiilibrium iterations 
in the simulation (in a static situation, nothing moves anymore; in a 
stationary simulation the averages are nearly constant since changes in one
direction are mostly cancelled by changes of other elements in the opposite
direction. There are many choices, one needs not mathematics, but mathematical 
thinking: precise and step-by-step.
 
\subsection{Warning against Mean Field Approximation}

\noindent
{\em This subsection contains many formulas and can be skipped}.
If you want to get answers by paper and pencil, you can use the mean field
approximation (also called molecular field approximation), which in economics
correponds to the approximation by representative agent. Approximate in the
first sum of Eq.(3) the $S_j$ by its average value, which is just the normalised
magnetisation $m = M/L^2 = \sum_i S_i/L^2$. Then the energy is
$$E = -J \sum_{<ij>} S_i m - H \sum_i S_i = -H_{eff} \sum S_i$$
with the effective field 

$$H_{eff} = H + \sum_j m = H + q m $$ 
where the latter sum runs over the $q$ neighbours only and is proportional to 
the magnetization $m$. Thus the energy $E_i$ of spin $i$ no longer is coupled 
to other spins $j$ and equals $\pm H_{eff}$. The probabilities $p$ for up and 
down orientations, according to Eq.(1), are now

$$p(S_i=+1) = \frac{1}{Z} \exp(H_{eff}/T) ; \quad p(S_i=-1) = \frac{1}{Z} 
\exp(-H_{eff}/T) $$ 
and thus

$$ m = p(S_i=+1) -  p(S_i=-1) = \tanh(H_{eff}/T) = \tanh[(H + q m)/T]$$
with the function tanh$(x) = (e^x - e^{-x})/(e^x+ e^{-x})$. This implicit
equation can be solved graphically; for small $m$ and $H/T$, tanh$(x) = x-x^3/3 +
\dots$
gives 

$$H/T = (1-T_c/T)m + \frac{1}{3} m^3 + \dots ; \quad T_c = qJ$$
related to Lev Davidovich Landau's theory of 1937 for critical phenomena 
($T$ near $T_c$, $m$ and $H/T$ small) near phase transitions. 

All this looks very nice except that it is wrong: In the one-dimensional
Ising model,  $T_c$ is zero
instead of the mean field approximation $T_c=qJ$. The larger the number of 
neighbours and the dimensionality of the lattice is, the more accurate is 
the mean field approximation. Basically, the approximation to replace $S_iS_j$
by an average $S_im$ takes into account the influenve of $S_j$ on $S_i$ but not
the fact that this $S_i$ again influences $S_j$ creating a feedback.
If the mathematics of this subsection looks 
deterrent, just ignore it; you are recommended to use computer simulations of 
single interacting spins, and not mean field theories. Outside of physics such 
simulations are often called ``agent based'' [Billari et al 2006, Bonabeau 
2002]; presumably the first one was the Metropolis algorithm
published in 1953 by the group of Edward Teller, who is historically known from 
the US hydrogen bomb and Strategic Defense Initiative (Star Wars, SDI). 

\section{Applications}

\subsection{Schelling Model for Social Segregation}

\noindent
Economics nobel laureate Thomas C. Schelling advised the US government on war
and peace in the 1960s and published a methodologically crucial paper (cited 
more than 500 times a decade later) [Schelling 1971, Fossett 2011, Henry et al 
2011] which introduced methods of statistical physics to
sociology. Each Ising spin corresponds to one of two ethnic groups (black and
white in big US cities) both of which prefer not to be surrounded by the other
group. The above Ising model then shows that for $T < T_c$ segregation 
emerges without any outside control: The simulated region becomes 
mostly white or mostly black. Unfortunately, Schelling simulated a more 
complicated version of the Ising lattice at $T=0$ which did not give large  
``ghettos'' like Harlem in New York City, but only small clusters of 
predominantly white and black residences. Only by additional randomness
[Jones 1985] (cited only 8 times, mostly by physicists) can ``infinitely''
large ghettos appear. It is easier to just simulate the above Ising model
[Sumour et al 2008,2011], 
which also allows people of one group to move into another city (or away
from the simulated region) to be replaced by residents of the other group. 
Sumour et al also cite earlier Schelling-type simulations by physicists.
Fore nearly three decades physicists ignored the Schelling model; now the
sociologists ignore the physics simulations of the last decade and the much
earlier Jones [1985] paper.  

As examples we give two iIsing-model figures from M\"uller et al [2008] where 
people also increase their amount $T$ of tolerance = social temperature if they 
see that their whole neighbourhood belongs to the same group as they themselves.
And afterwards they slowly forget and reduce their tolerance. With changing
frogeting rate one may either observe lots of small clusters, Fig.1, or one
big ghetto, Fig.2. 

\begin{figure}[hbt]
\begin{center}
\includegraphics[angle=-90,scale=0.5]{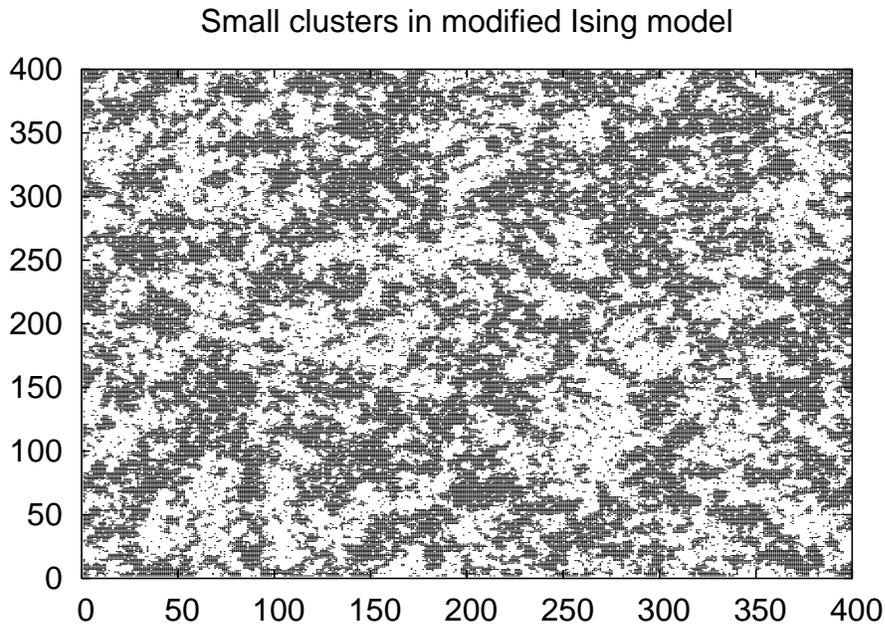}
\end{center}
\caption{Cluster formation for slow forgetting in the Ising modification of
M\"uller et al [2008]; no large ghetto is formed.
}
\end{figure}

\begin{figure}[hbt]
\begin{center}
\includegraphics[angle=-90,scale=0.5]{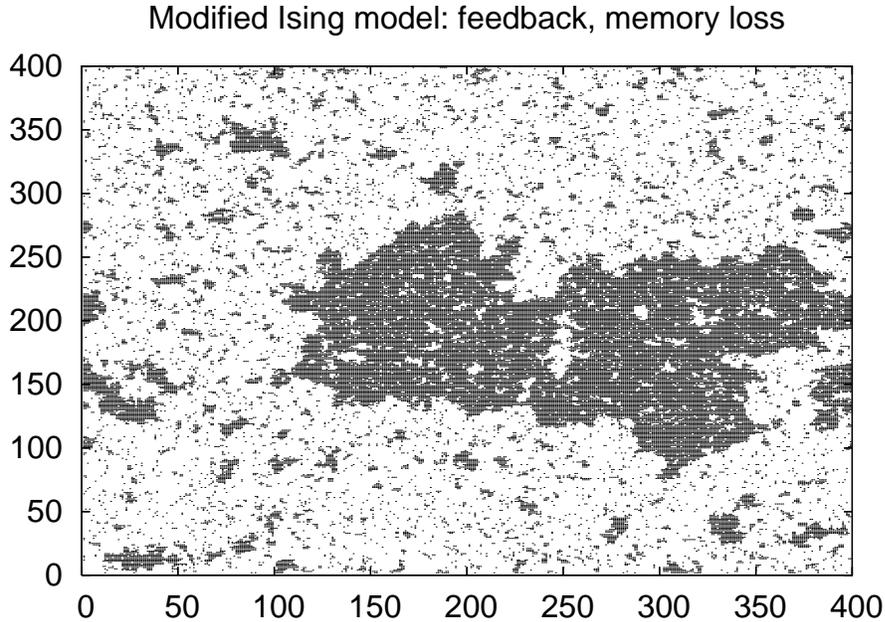}
\end{center}
\caption{In the same model of the previous figure, a large ghetto is formed if
people forget faster their learned tolerance.
}
\end{figure}

About simultaneously with this Schelling paper, physicist Weidlich started his 
sociodynamics approach to apply the style of physics to social questions
[Weidlich 2000], see also [Galam 2008]. 

\subsection{Sociophysics and Networks}

\noindent
A good overall review of the sociophysics field (of which the Schelling model
is just one example) was given by Castellano et al [2009].
Of particular interest is the reproduction of universal properties of election
results with many candidates: The curves of how many candidates got $n$ votes 
each are similar to each other [Castellano 2009]. Car traffice [Chowdhury et 
al [2000], economic markets [Bouchaud and Potters 2010, Bonabeau 2002], 
opinion dynamics [Malarz et al 2011], stone-age culture [Shennan 2001], 
histophysics [Lam 2002], languages [Schulze et al 2008], Napoleon's decision 
before the battle of Waterloo [Mongin 2008], religion [Ausloos and Petroni 
2009], political secession from a state [Lustick 2011], demography on social 
networks [Fent et al. 2011], insurgent wars in Iraq and Afghanistan [Johnson
et al. 1944],
... are other applications. For example, non-physicists [Holman et al 2011]
have acknowledged that physicists Serva and Petroni (2008) may have been the 
first to use Levinstein distances to calculate the ages of language groups.
(These distances are differences between words for the same meaning in different
languages.) 

Students in class may sit on a square lattice, but normally humans do not. (In
universities, mostly only a part of the lattice is occupied, which physicists
simulate as a ``dilute'' square lattice.) They
may be connected by friendship or job not only with nearest neighbors but 
also with people further away. Such networks, investigated for a long time
by sociologists [Stegbauer and Haeussling 2010], 
were studied by physicists intensively for a dozen years [Barab\'asi 2002, 
Albert and Barab\'asi 2002, Bornholdt and Schuster 2003, Cohen and Havlin
2010]. In the Watts-Strogatz (or ``small world'') network a random fraction of
nearest-neighbour bonds
is replaced by bonds with sites further away, selected randomly from the whole
lattice. If that fraction approaches unity, one obtains the Erd\"os-R\'enyi
networks, a limit of percolation theory [Flory 1941]. More realistic are the 
scale-free Barab\'asi-Albert networks, where the network starts with a small 
core and then each newly added site forms a bond with a randomly selected 
already existing member of the network. The selection probability is 
proportional to the number of bonds which the old member already has acquired: 
Famous people get more attention and more ``friends'' than others; no lattice 
is assumed here anymore. In all these networks,
the average number of bonds needed to connect two randomly selected sites 
increases logarithmically with the number of sites in the network, whereas 
for $d$-dimensional lattices this average number of bonds increases stronger
with a power law, exponent $1/d$. Having simulated one network, one can also
study connected sets of networks or other social networks [Watts, Dodds,
Newman 2002], or demography on them [Fent et al 2011]. 

The latest application is Statistical Justice: In May 2011, John Demjanjuk 
was sentenced for having helped in 1943 in the murder of more than 28,000 
Dutch Jews in the Nazi concentration camp of Sobib\'or. One knows who was 
deported but not who survived the transport from the Netherlands to Poland.
And one does not know which duties the accused had there on which day. Thus 
all acts of the camp guards were regarded as having helped in their murder,
and the number near 28,000 was estimated from the average death rate during 
the transports. The verdict thus gave neither the name of a murder victim 
nor the day of a murder, but was entirely based on statistical averages.
[Times 2011]. 

\section{Appendix: How to Program the Ising Model}

The following Fortran manual and program are both short and should encourage 
the reader to learn this technique. 

\subsection{Fortran Manual}

\noindent
Fortran = formula translator an early language (above machine code or assembler)
for computer programming;
many others followed and in particular C$^{++}$ is widespread, but nevertheless 
this tutorial uses Fortran which is closer to plain English and allows to easily
find a typical programming error (using an array outside its defined bounds). 
If your Fortran program is called {\tt name.f}, it can be compiled with {\tt 
f95 -O name.f} (or 77 instead of 95; {\tt O} = optimisation), and executed with 
{\tt ./a.out} (or just {\tt a.out}). The just mentioned error message appears 
when using {\tt f95 -fbounds-check name.f ; ./a.out}, but execution then is much
slower and thus instead {\tt -O} should be used after error correction.

Fortran commands usually start in column 7 and end before column 73. A C in
column 1 signifies a comment for the reader, to be ignored by the computer.
In column 6 we write a 1 if this line is a continuation of the previous line,
while columns 2 to 5 are reserved for labels, i.e. numbers to control the flow 
of commands. For example, {\tt GOTO 7} means to jump to the line labelled by 7.

Variable names start with a letter; names starting with I, J, K, L, M, N signify
integers without rounding errors, other names are real (floating-point) numbers
and nearly always have rounding errors. The operations $+, -, *, /$ and
{\tt SQRT, COS, SIN, EXP} etc have their usual meaning except that {\tt N/M} is
always rounded downwards to an integer value; e.g. 3/5 is zero. Also {\tt
I = X} means rounding downwards. Since the natural logarithm is not an integer 
it is denoted by {\tt ALOG} instead of log.

Decisions are made automatically, e.g.

\quad \quad {\tt IF(A.GT.0) B = SQRT(A)}

\noindent
where {\rm .GT.} means greater than, with analogous meanings for {\tt .LT.,
.GE, .LE., EQ., .NE., .NOT., .AND., .OR.} \quad .

A loop is executed by

\quad \quad {\tt DO 99 K=M,N}

\noindent
which means that all lines from this line down to and including the line with
label 99 are executed for $k=m, m+1, m+2, ..., n$. One may put inner loops
into outer loops, if needed.

Arrays need to be declared at the beginning of a program, for example through

\quad \quad {\tt DIMENSION A(100,100), B(100), C(L)}

\noindent
Here, if $C$ has an arbitrary dimension $L$, then $L$ must be given a value 
before this dimension statement through

\quad \quad {\tt PARAMETER(L = 100) }
 
\noindent
and must not be changed throughout the program. Similarly, variables can 
be initialised via a data line like

\quad \quad {\tt DATA L/100/, B/100*1.0/}

\noindent
but only once at the beginning of the program, not later again.

Results are best printed out through 

\quad \quad {\tt PRINT *, x, y, z}  

\noindent
Thereafter execution should stop with a
{\tt STOP} line, followed by an {\tt END} line.

The statement 

\quad \quad {\tt n = n+1}

\noindent 
is not an equality (which then could be simplified to the nonsensical 0 = 1)
but a command to the computer: to find the place in the  memory where the 
variable {\tt n} is stored, to get the value of {\tt n} from there, to add
one to it, and to store the sum in that same memory place as the new value 
for {\tt n}. Some computer languages therefore use := instead of the simpler
but misleading = sign. 

Normally it does not matter whether or not CAPITAL letters are used. 
The computer language Basic is rather similar to Fortran. Now we bring
a complete program to simulate the Ising model on the square lattice.

\subsection{Ising Model Program}

\noindent
{\small
\begin{verbatim}
c     heat bath 2D Ising in a field
      parameter(L=1001,Lmax=(L+2)*L)
      dimension is(Lmax),ex(-4:4)
      data t,mcstep,iseed/0.90,1000,1/,h/+0.50/,ex/9*0.0/
      print *, '#', L,mcstep,iseed,t,h
      x=rand(iseed)
      Lp1=L+1
      LspL=L*L+L
      L2p1=2*L+1
      do 1 i=1,Lmax
 1      is(i)=1
      do 2 ie=-4,4,2
        x=exp(-ie*2.0*0.4406868/t-h)
 2      ex(ie)=x/(1.0+x)
      do 3 mc=1,mcstep
        mag=0
        do 4 i=Lp1,LspL
c         if(i.ne.L2p1) goto 6
c         do 5 j=1,L
c5          is(j+LspL)=is(j+L)
 6        ie=is(i-1)+is(i+1)+is(i-L)+is(i+L)
          is(i)=1
          if(rand().lt.ex(ie)) is(i)=-1
 4        mag=mag+is(i)
c       do 7 i=1,L
c7        is(i)=is(i+L*L)
 3      if(mc.eq.(mc/100)*100) print *, mc, mag
      stop
      end
\end{verbatim}
}
 
The parameter line fixes the size of the $L \times L$ square; the sites in it
are numbered by one index, typewriter style. Thus the right neighbor of 
site $2L$ is $2L+1$ and sits on the left end of the next line: Helical 
boundary conditions. The lower neighbour of site $2L$ is $2L+L$, the upper
neighbour is $2L-L$, and the left neighbor is $2L-1$. The spins in the top and 
bottom buffer lines ($1 \dots L$ and $L^2+L+1 \dots L^2+2L)$ stay in their
initial up orientation; if instead one wants periodic boundary conditions
in the vertical direction (better to reduce boundary influence), 
one has to omit the five comment symbols C at and before loops 5 and 7.

The temperature enters the data line in units of $T_c$, i.e. $T = 0.9 T_c$ 
in this example; the known value $J/T_c = 0.44068\dots$ is used nine lines
later. In the same line the field, in units of $k_BT$, is given as 0.5.
In the line after the first print statement, {\tt rand(iseed)} 
initialises the random number generator (see next subsection for warning
and improvement); a different seed integer gives different random numbers. Later
{\tt rand()} produces the next ``random'' number between 0 and 1 from the last 
one in a reproducible but hardly predictable way. Loop 2 determines the Boltzmann 
probabilities {\tt iex} needed for Eqs.(1) and finishes the initialisation.
Loop 3 makes {\tt mcstep} iterations (time steps = Monte Carlo steps per spin).
Loop 4 runs over all spins except those in the two buffer lines and determines
the local interaction energy {\tt ie} as the sum over the four neighbour spins.
Then we set the spin to $+1$, and if the conditions of Eqs(1) so require,
instead it is set to $-1$. (Generally, a command is executed with a probability 
$p$ if the random number {\tt rand()} is smaller than $p$.) We print out the
magnetisation only every hundred time steps in order to avoid too much data
on the computer screen. The results are plotted in Fig.3 and show that the
temperature is very low: Even though it is only 10 percent below the critical
temperature $T_c$, the magnetisatuion barely changes from its initial value
1002001 and is after 100 iterations already in equilibrium, also due to the 
applied field. In three instead of two dimensions, without a field, and at
temperatures closer to $T_c$ longer times are needed for equilibration.

\begin{figure}[hbt]
\begin{center}
\includegraphics[angle=-90,scale=0.34]{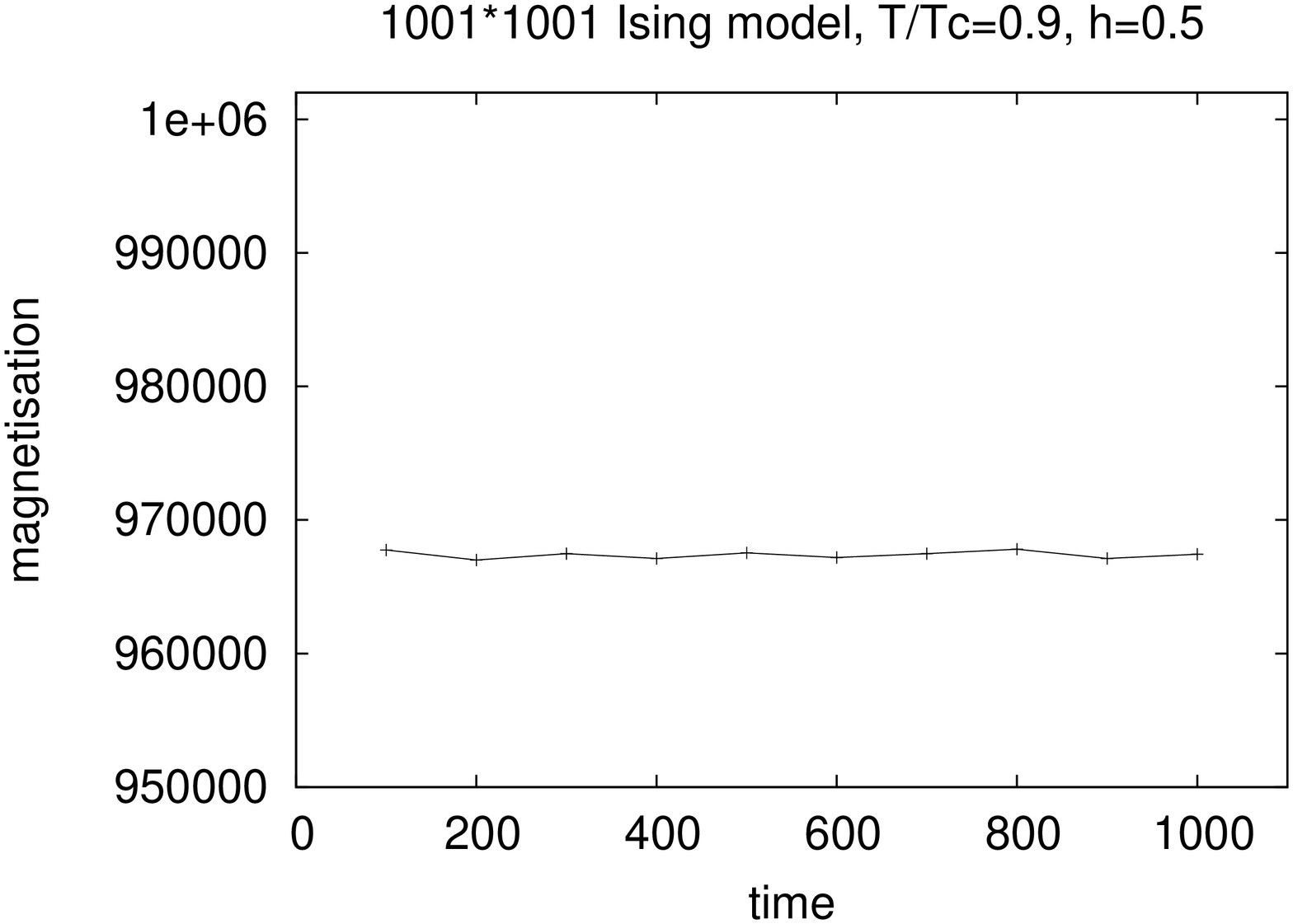}
\includegraphics[angle=-90,scale=0.34]{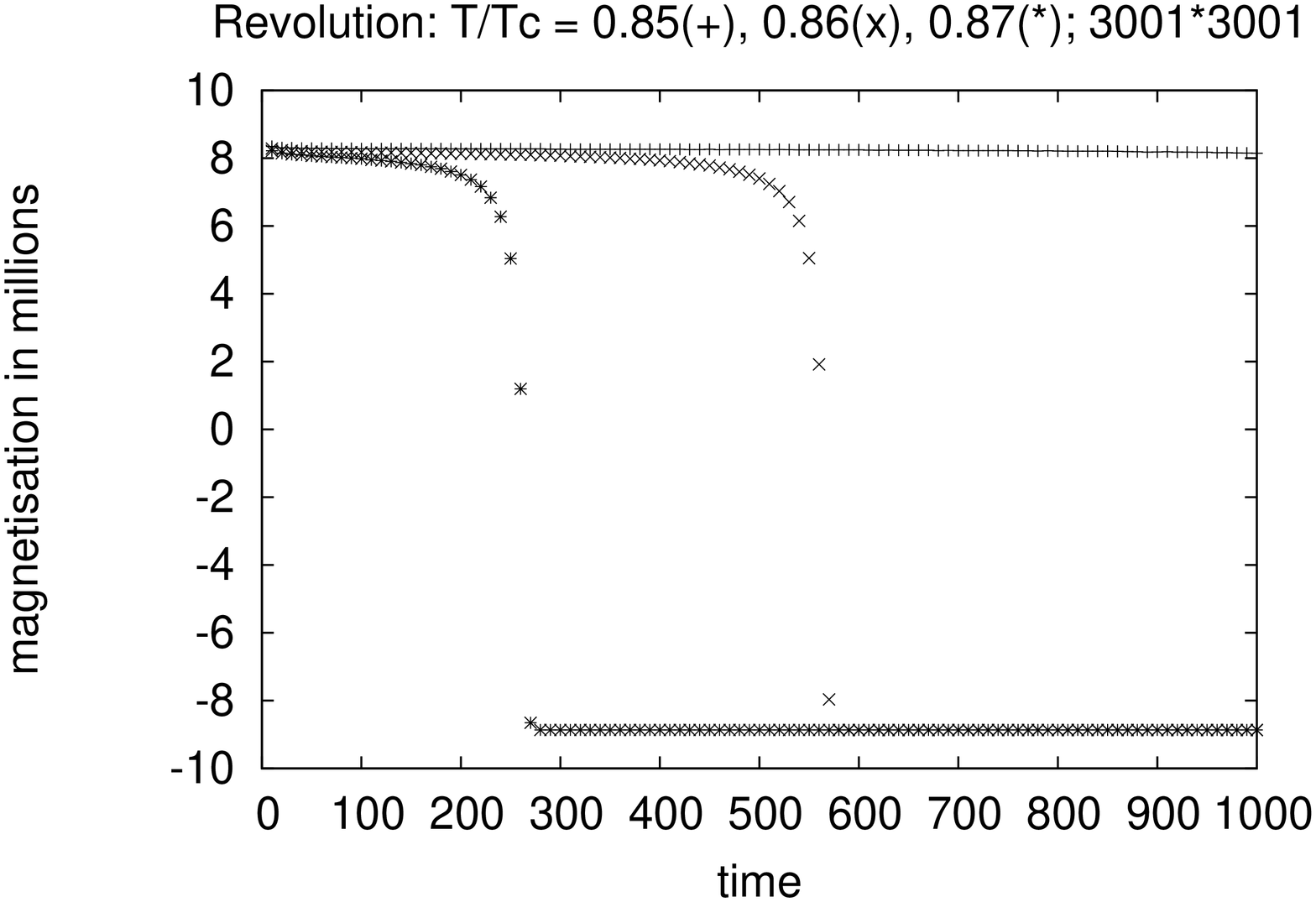}
\end{center}
\caption{Top: Results from the listed Ising program. Bottom: Revolutions
}
\end{figure}

Perhaps you find it more interesting to simulate revolutions, as in Fig.1b,
where the field $h$ was equal to the fraction of overturned spins (suggestion
of Sorin Solomon for Bornholdt-type model).  
So, what is difficult about computations? Do you know a shorter Fortran manual?

\subsection{Random Numbers}

\noindent
The above {\tt rand} produces random numbers in an easily programmed way, but
often this may be slow and/or bad, or the used algorithm is unknown to the
user. It is better to program random number generation explicitely. If you
multiply by hand two nine-digit integers, you may easily predict the first and 
the last digit of the product, but hardly the digits in the middle, except by
tediously doing the whole multiplication correctly. Similarly, if {\tt ibm} is
a 32-bit odd integer, then the product 

\quad \quad {\tt ibm=ibm*16807}

\noindent
is again an odd integer, and normally requires 46 bits. (A bit = binary digit
is a zero or one in a computer.) The computer throws 
away the leading bits and leaves the least significant 32 bits. The first bit 
gives the sign, thus plus times plus gives minus in about half the cases, in
contrast to what you learned in elementary school. The last of these 
remaining bits is predictably always set to one (odd integers) but the 
leading (most significant) bits are quite random. (Actually, your computer
may do something very similar when you call {\tt rand()}.) More precisely,
they are pseudo-random; in order to search for errors one wants to get exactly
the same random numbers when one repeats a simulation with the same seed.

These random 32-bit integers {\tt ibm} between --2147483647 and $+2147483647 = 
2^{31}-1$
can be transformed into real numbers through {\tt ran=factor*ibm+0.5} where
{\tt factor = 0.5/2147483647}, but it is more efficient to normalize the 
propabilities $p$ (here {\tt = ex/(1.0+ex)} to the full interval of 32-bit
integers through {\tt (2.0*p-1.0)*2147483647}, once at the beginning of the
simulation. Then the next random integers
{\tt ibm} simply have to be compared with this normalized probability. In the
above Ising program, one then stores the Boltzmann probabilities as {\tt 
dimension iex(-4:4)} at the beginning, calculates them through

\noindent
\quad {\tt 2 \quad iex(ie)=(2.0*ex/(1.0+ex) - 1.0)*2147483647}

\noindent
in the initialisation, and later merely needs

\quad \quad {\tt ibm=ibm*16807}

\quad \quad {\tt if(ibm.lt.iex(ie)) is(i)=-1}

\noindent
in the above Ising program. 

With 32 bits the pseudo-random integers are repeated after $2^{29}$ 
such multiplications with 16807, which is a rather small number for 
today's personal computers. It is better to use 64 bits via

\quad \quad {\tt integer*8 ibm, iex} 

\noindent
at the beginning of the program
in order to get many more different random numbers, using e.g.

\noindent
\quad {\tt 2  \quad iex(ie)=2147483647*(4.0*ex/(1.0+ex) - 2.0)*2147483647}

\noindent
for the normalized 64-bit probabilities. Now the quality is much better without
much loss in speed; unfortunately one now can make much more programming
errors involving these random numbers. 

\bigskip
\subsection{History}

Now comes a list of extended abstracts, a few pages each, about what this
author finds interesting in recent history, available on request from 
dstauff@thp.uni-koeln.de: 
\medskip

\noindent
 Who is to blame for World War I \\
 No miracle on the Marne, 9/9/1914 \\
 Lies and Art. 231 of Versailles Peace Treaty 1919 \\
 Was Hitler's 1941 attack against the Soviet Union a preemptive war ? \\
 Had Hitler nearly gotten Mosnow in 1941 ? \\
 Why was there no joint Japanese attack when Hitler attacked the Soviet Union \\
 The Sea Battle of Leyte, 25 October 1944 \\
 Did Soviet tanks approach Tehran in March 1946 ? \\
 Missed chance for peace in Korea, October 1950? \\
 Stalin's proposal of March 1952 for a united Germany \\
 1956: West German finger on the nuclear trigger ? \\
 Tank confrontation at Checkpoint Charlie 10/1961 \\
 The 1962 Cuban Missile crisis: security or prestige? \\
 Lyndon B. Johnson (1908-1973) and the Dominican crisis (US Invasion 1965) \\
 1990: East Germany into NATO ? \\
 Kosovo War 1999  \\
 The start of the Libyan war, March 2011 \\

Thanks to S. Wichmann, T. Hadzibeganovic, M. Ausloos, and T. Fent for a 
critical reading of the manuscript. 

\bigskip
\bigskip

\parindent 0pt

Albert R. and Barab\'asi A.L. [2002], ``Statistical mechanics of complex 
cetworks'', {\it Reviews of Modern Physics} {\bf 74}, 47-97.

\medskip
Ausloos M. and Petroni F., ``Statistical dynamics of religion evolutions'',
{\it Physica A} {\bf 388}, 4438-4444; M. Ausloos [2010], ``On religion and 
language evolutions seen through mathematical and agent based models'', in 
{\it Proceedings of the First Interdisciplinary CHESS
Interactions Conference},  C. Rangacharyulu and E. Haven, Eds., World
Scientific, Singapore, pp. 157-182. 

\medskip
Barab\'asi A.L. [2002] ``Linked'', Perseus, Cambridge.

\medskip
Billari F.C., Fent T., Prskawetz A., and Scheffran J. [2006] {\it
Agent-based computational modelling}, Physica-Verlag, Heidelberg.

\medskip
Bonabeau E. [2002] ``Agent-based modelling: Methods and techniques for 
simulating human systems'', Proc. Natl. Acad. Sci. USA {\bf 99}, 7280-7287.

\medskip
Bornholdt S. and Schuster H.G. [2003] {\it Handbook of graphs and networks},
Wiley-VCH, Weinheim.

\medskip
Bouchaud J.P. and Potters M. [2009] {\it Theory of financial risks and 
derivative pricing}, Cambridge University Press, Cambridge. 

\medskip 
Bridson R. and Batty C [2010] ``Computational physics in film'', {\it Science}
{\bf 330}, 1756-1757.
 
\medskip
Castellano C., Fortunato S., Loreto V., [2009] ``Statistical physics of social
dynamics'', {\it Rev. Mod. Physics} {\bf 81}, 591-646.

\medskip
Chowdhury D., Santen L., Schadschneider A. [2000] ``Statistical physics of 
vehicular traffic and some related systems'', {\it Physics Reports} {\bf 329},
199-329. 

\medskip
Cohen R. and Havlin S. [2010] {\it Complex Networks}, Cambridge University
Press, Cambridge.

\medskip
Diamond, J [1997] {\it Guns, Germs, and Steel}, Norton, New York. 

\medskip
Fent T., Diaz B.A, Prskawetz A. [2011] ``Family policies in the context of 
low fertility and social structure'', {\it Vienna Inst. Demogr. Working Paper}
2/2011 (www.oeaw.at/vid).

\medskip
Flory, P.J. [1941] ``Molecular size distribution in three-dimensional 
polymers: I, II, III'', {\it J. Am. Chem. Soc.} {\bf 63}, 3083, 3091, 3096.

\medskip
Fossett, M. [2011] ``Generative models of segregation'', {\it J. Math. 
Sociology} {\bf 35}, 114-145.

\medskip
Galam S. [2008] ``Sociophysics: A review of Galam models'', {\it Int. J. Mod. 
Phys. C} {\bf 19}, 409-440.

\medskip
Geiss, I. [2008] {\it Geschichte im \"Uberblick}, Anaconda, K\"oln.

\medskip
Henry, A.D., Pralat, P., Zhang, C-Q. [2011] Proc. Natl. Acad. Sci. USA 108,
8505-8610.

\medskip
Holman, E.W., 14 coauthors [2011], ``Automated dating of the world's language 
families based on lexical similarity'', preprint for {\it Current Anthropology}.

\medskip
Johnson, N., Carran, S., Botner, J., Fontaine, K., Laxague, N., Nuetzel, P., 
Turnley, J., Tivnan, B. [2011] ``Pattern in escalations in insurgent and
terrorist activity'', {\it Science} {\bf 333}, 81-84.

\medskip
Jones F.L. [1985] ``Segregation models of group segregation'',  {\it Aust. New 
Zeal. J. Sociol.} {\bf 21}, 431-444.

\medskip
Lim M., Metzler R., Bar-Yam Y. [2007] ``Global pattern formation and 
ethnic/cultural violence'', {\it Science} {\bf 317}, 1540-1544. See 
also Hadzibeganovic T. et al [2008], Physica A {\bf 387}, 3242-3252. 

\medskip
Lustick J. [2011] ``Secession of the center: A virtual probe of the prospects 
for Punjabi secessionism in Pakistan and the Secession of Punjabistan'', 
{\it Journal of Artificial Societies and Social Simulation} {\bf 14}, issue 1, 
paper 7 (electronic only via jasss.soc.surrey.ac.uk).

\medskip
Malarz K., Gronek P. and Ku{\l}akowksi K. [2011] ``Zaller-Deffuant model of mass
opinion'', {\it Journal of Artificial Societies and Social Simulation} {\bf 14},
issue 1, paper 2 (electronic only via jasss.soc.surrey.ac.uk).

\medskip
Mongin P. [2008] ``Retour \`a Waterloo - Histoire militaire et th\'eorie des 
jeux'', {\it Annales. Histoire, Sciences Sociales} {\bf 63}, 39-69.

\medskip
M\"uller, K., Schulze, C., Stauffer, D. [2008] ``Inhomogeneous and 
self-organized temperature in Schelling-Ising model'', {\it Int. J. Mod. Phys. 
C} {\bf 19}, 385-391. 

\medskip
Prasos, J. [2009] ``Vietnam'', University Press of Kansas, Lawrence 2009, p.
xiii.
 
\medskip
Richardson L.F. [1935] ``Mathematical psychology of war'', {\it Nature} {\bf 
135}, 830-831 and {\bf 136}, 1025-1026.

\medskip
Schelling T.C. ``Dynamic models of segregation'' [1971] {\it J. Math. Sociol.}
{\bf 1} 143-186.

\medskip
Schulze C., Stauffer D., and Wichmann S. [2008] ``Birth, survival and death of 
languages by Monte Carlo simulation''. {\it Comm. Comput. Phys.} {\bf 3}, 
271-294.

\medskip
Shennan S. [2001] ``Demography and cultural innovation: a model and its 
implications for the emergence of modern human culture'', {\it Cambridge 
Archeol. J.} {\bf 11}, 5-16.

\medskip
Stauffer D., Moss de Oliveira S., de Oliveira P.M.C., S\'a Martins J.S. 
[2006]. {\it Biology, sociology, geology by computational physicists}, 
Elsevier, Amsterdam.

\medskip
Stegbauer C. and Haeussling R. (eds.) [2010], {\it Handbuch Netzwerkforschung},
VS-Verlag, Wiesbaden.

\medskip
Sumour M.A., El-Astal, A.H., Radwan M.M., Shabat, M.M. [2008],
Urban segregation with cheap and expensive residences,
{\it Int. J. Mod. Phys. C} {\bf 19}, 637-645.  

\medskip
Sumour M.A., Radwan M.M., Shabat, M.M. [2011], ``Highly nonlinear Ising model
and social segregation'',  arXiv:1106.5574 (electronically only on arXiv.org
section physics).

\medskip
Times: nytimes.com May 12, 2011 ``Demjanjuk''.

\medskip
Vitanov, N.K. and Ausloos, M. R.[2011] `` Knowledge epidemics and population 
dynamics models for describing idea diffusion''. In: 
{\it Models of Science Dynamics - Encouters between Complexity
Theory and Information Sciences}, ed. by Scharnhorst, A. Boerner, K.,
P. van den Besselaar. Springer, Berlin Heidelberg (forthcoming)

\medskip
Watts, D.J., Dodds, P.S., and Newman, M.E.J. [2002] ``Identity and search in 
social networks'', {\it Science} {\bf 296},1302-1305.

\medskip
Weidlich W.  [2000] {\it Sociodynamics; a systematic approach
to mathematical modelling in the social sciences}  Harwood Academic Publishers;
2006 reprint: Dover, Mineola (New York).

%\newpage
\bigskip
\bigskip

\noindent
Dietrich Stauffer is retired professor of theoretical physics and studies 
history (mostly 20th century, mostly diplomatic) since retirement. Before that
he worked on Monte Carlo simulations, like Ising models, percolation, ageing,
opinion dynamics. 

\bigskip
\noindent
Institute for Theoretical Physics, Cologne University,
D-50923 K\"oln, Euroland
\end{document}